# Efficient Up-Conversion in CsPbBr$_3$ Nanocrystals via Phonon-Driven Exciton-Polaron Formation

*Abdullah S. Abbas[1,†], Beiye C. Li[4,6,7], Richard D. Schaller[8,9], Vitali B. Prakapenka[10], Stella Chariton[10], Gregory S. Engel[4,5,6,7], A. Paul Alivisatos[1,2,3,4,5,6, §,*]*

[1]Department of Materials Science and Engineering, University of California, Berkeley, Berkeley, California 94720, United States

[2]Department of Chemistry, University of California, Berkeley, Berkeley, California 94720, United States

[3]Materials Sciences Division, Lawrence Berkeley National Laboratory, Berkeley, California 94720, United States

[4]Department of Chemistry, The University of Chicago, Chicago, Illinois 60637, United States

[5]Pritzker School of Molecular Engineering, The University of Chicago, Chicago, Illinois 60637, United States

[6]James Franck Institute, The University of Chicago, Chicago, Illinois 60637, United States

[7]Institute for Biophysical Dynamics, The University of Chicago, Chicago, Illinois 60637, United States

[8]Department of Chemistry, Northwestern University, Evanston, Illinois 60208, United States

[9]Center for Nanoscale Materials, Argonne National Laboratory, Lemont, Illinois 60439, United States

[10]Center for Advanced Radiation Sources, The University of Chicago, Chicago, Illinois 60637, United States

Present addresses:

[†]Department of Chemistry, The University of Chicago, Chicago, Illinois 60637, United States

[§]Department of Chemistry and Pritzker School of Molecular Engineering, The University of Chicago, Chicago, Illinois 60637, United States

*To whom correspondence may be addressed: paul.alivisatos@uchicago.edu

ABSTRACT

Lead halide perovskite nanocrystals demonstrate efficient up-conversion, although the precise mechanism remains a subject of active research. This study utilizes steady-state and time-resolved spectroscopy methods to unravel the mechanism driving the up-conversion process in CsPbBr$_3$ nanocrystals. Employing above- and below-gap photoluminescence measurements, we extract a distinct phonon mode with an energy of ~7 meV and identify the Pb-Br-Pb bending mode as the phonon involved in the up-conversion process. This result was corroborated by Raman spectroscopy. We confirm an up-conversion efficiency reaching up to 75%. Transient absorption measurements under conditions of sub-gap excitation also unexpectedly reveal coherent phonons for the subset of nanocrystals undergoing up-conversion. This coherence implies that the up-conversion and subsequent relaxation is accompanied by a synchronized and phased lattice motion. This study reveals that efficient up-conversion in CsPbBr$_3$ nanocrystals is powered by a unique interplay between the soft lattice structure, phonons, and excited states dynamics.

KEYWORDS: Perovskite Nanocrystals, Up-Conversion, Coherent Phonon, Polaron, Anti-Stokes Photoluminescence.

Lead halide perovskite nanocrystals have emerged as a highly promising semiconducting nanomaterial for phonon-driven up-conversion, owing to their ease of synthesis and near-unity photoluminescence quantum yield (**PLQY**) [1–5]. In phonon-driven up-conversion, a phonon is annihilated upon sub-bandgap light absorption to arrive at an excited state population followed by anti-Stokes photoluminescence (**ASPL**), where the emitted photon energy is higher than that of the excitation photon (**Scheme 1**). Once an electron undergoes up-conversion to arrive at the conduction band edge, radiative emission without generation of a phonon leads to an overall reduction in phonon population, while non-radiative recombination results in phonon generation.

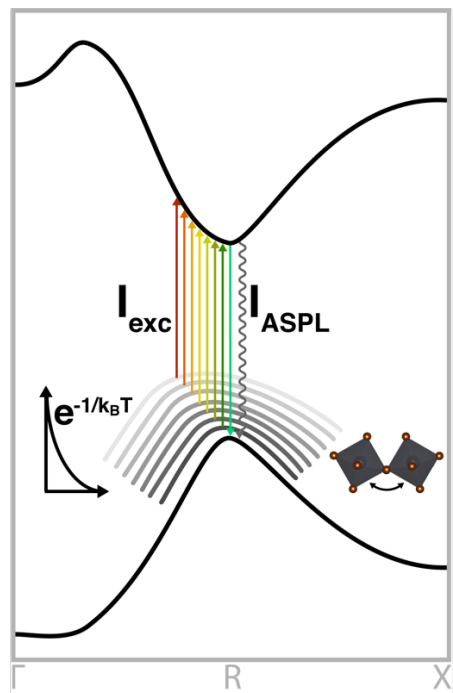

**Scheme 1:** Energy band diagram of $CsPbBr_3$ nanocrystals[6] illustrating the up-conversion process and anti-Stokes photoluminescence (ASPL). Electronic states form due to lattice vibrations and are thermally populated, for example with bending mode phonons[7]. Sub-bandgap phonon-assisted

light absorption promotes a valence electron to the conduction band and subsequent ASPL decreases the phonon population.

Among promising semiconducting nanomaterials candidates for up-conversion[4], cesium lead tribromide (**CsPbBr$_3$**) perovskite nanocrystals exhibit substantial up-conversion efficiencies[1,4,5], as well as a remarkably low trap density and high tolerance for shallow surface defects, which contribute to their near-unity PLQY[8–10]. On an ensemble and single-particle level, CsPbBr$_3$ nanocrystals can up-convert with efficiencies up to 75% when the excitation source is detuned below the gap by 23 meV[5]. However, the phonon-assisted up-conversion process is particularly delicate and complex because more vibrational degrees of freedom create a larger hot thermal population to excite, but these same degrees of freedom tend to abrogate photoluminescence in favor of non-radiative relaxation. Additionally, coherent phonon oscillations have been directly observed in layered perovskites[11,12], and a recent report on quasi-two-dimensional perovskites showed that the coherent oscillations are involved in the up-conversion process of carriers[13].

Here, we deploy a combination of steady-state and excited-state measurements to resolve the specific phonons involved in the efficient up-conversion process within CsPbBr$_3$ nanocrystals, determine up-conversion efficiency at different detuning energies, and investigate how electron-phonon coupling changes in nanocrystals undergoing up-conversion. Using a tunable laser, we leverage the exponential relationship between the excitation wavelengths and the ratio of laser intensity to ASPL intensity to extract the phonon energy related to the up-conversion process. We identify the Pb-Br-Pb bending mode, with energy ~ 7 meV[7,14–16], as the phonon involved in up-conversion within CsPbBr$_3$ nanocrystals. In excited-state measurements, the above-gap excitation results in the expected intraband cooling characteristics. However, with below-gap excitation, an unexpected coherent phonon oscillations emerges at early times as exciton-polarons form within the subset of nanocrystals undergoing up-conversion. These coherent lattice vibrations are synchronized to the frequency of the Pb-Br-Pb bending phonon mode, which implies that the up-conversion and subsequent relaxation is accompanied by a synchronized and phased lattice motion. In all, this work provides invaluable insight into how the intricate interplay between the soft lattice structure, phonons, and excited states dynamics contributes to efficient up-conversion in CsPbBr$_3$ nanocrystals.

# RESULTS & DISCUSSION

## Steady-State Dynamics

We first set out to identify the specific phonon modes involved in up-conversion in CsPbBr$_3$ nanocrystals. We use high-quality nanocrystals to avoid non-radiative pathways that might otherwise obscure the sub-bandgap dynamics involved in the up-conversion process[17]. CsPbBr$_3$ nanocrystals were synthesized following a modified version of previous reports[1,18]. **Fig. 1a** shows the steady-state absorption and photoluminescence (**PL**) spectra with band edge emission peak at ~ 2.435 eV and a full width at half maximum (**FWHM**) of 81.38 meV. The as-synthesized (no-post synthetic treatment) CsPbBr$_3$ nanocrystals exhibit very high absolute PLQY, approaching ~100% unity within 2.0% uncertainty, as shown in **Fig. 1b**. This near-unity PLQY, an attribute of lead-halide perovskite nanocrystals[2,3,19], highlights their suitability for our study. The nanocrystals are highly monodispersed, **Fig. 1c**, which is also indicated by the narrow linewidth in the PL spectrum.

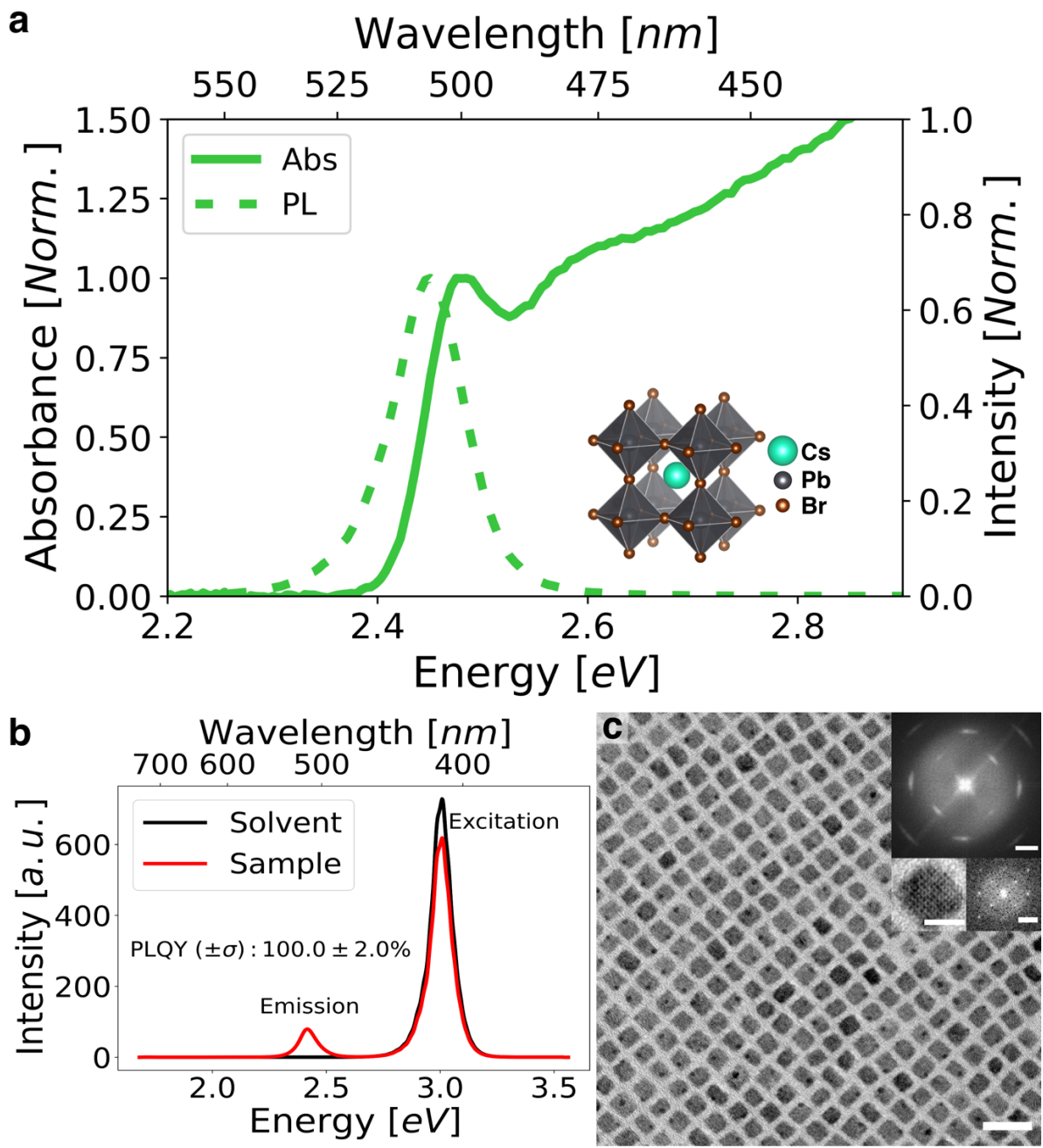

**Figure 1: Optical and structural characterizations of CsPbBr$_3$ nanocrystals. a**, steady-state absorption and photoluminescence (PL) spectra, with inset illustrating the crystal structure made with Vesta[20]. **b**, Near-unity photoluminescence quantum yield (PLQY) measurement. **c**, Transmission electron microscopy (TEM) image shows highly monodispersed nanocrystals and top inset shows the Fast Fourier Transform (FFT) (scale bars 20 nm and 1 nm$^{-1}$, respectively).

Bottom insets show high-resolution TEM image of a single nanocrystal with FFT (scale bars 5 nm and 2 nm$^{-1}$, respectively).

We employed a tunable laser with a narrow bandwidth of 3 nm to photoexcite the high-quality nanocrystals below the band edge at different wavelengths. A long-pass filter with a wavelength of 515 nm was employed to prevent any unexpected broad bandwidth in the laser spectrum from exciting the sample, as depicted in Fig. S1. **Fig. 2a** (and inset) shows the ASPL as a result of excitation below the band edge at room temperature over a wavelength range of 515 to 550 nm. The ASPL in CsPbBr$_3$ is expected[5], and we find that with increasing excitation wavelengths, the ratio of laser intensity to the ASPL intensity follows an exponential form (**Fig. 2b**, see also log plot in the inset). We fit the data to an Arrhenius equation to extract a characteristic energy ($E_{ch}$) as follows:

$$\frac{I_{exc}}{I_{aspl}} \propto e^{-\left(\frac{E_g - h\nu}{E_{ch}}\right)}$$

where $E_g$ is the bandgap energy, $h\nu$ is the excitation energy, and $E_{ch}$ is a characteristic energy. This characteristic energy involves both $\kappa_B T$ at room temperature (RT) and the phonon energy ($E_p$) responsible for the up-conversion process. From the fit shown in **Fig. 2b**, we retrieve $E_p = 7.28$ meV ($58.68$ cm$^{-1}$) which corresponds to the Pb-Br-Pb bending mode in CsPbBr$_3$ nanocrystals[15,21]. The presence of this low frequency is confirmed by Raman spectroscopy (peak $62.98$ cm$^{-1}$), as shown in **Fig. 2c** (see Fig. S2 for raw data and background fit). The lowest observed frequency ($25.42$ cm$^{-1}$) corresponds to the octahedral twist[22,23]. The $104.66$ and $127.38$ cm$^{-1}$ frequencies are assigned to the longitudinal optical (LO) phonon modes associated with the stretching of the Pb-Br bonds in the [PbX$_3$]$^-$ sublattice, while the

309.88 and 317.62 cm$^{-1}$ modes are assigned to higher order scattering based on previous reports[7,15,16,21].

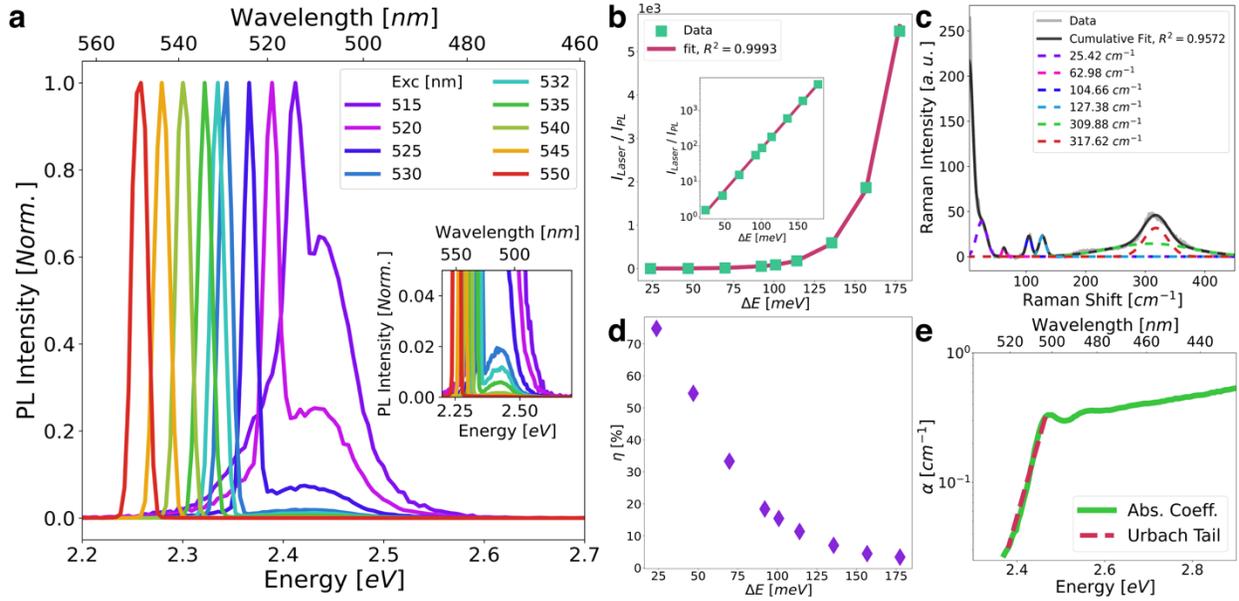

**Figure 2: a**, Below-gap photoexcitation shows distinct band edge emission characteristics. **b**, Analysis of the laser to PL intensity ratio reveals an exponential relationship. Inset, presented on a logarithmic scale, confirms a linear trend. **c**, Raman spectrum of CsPbBr$_3$ nanocrystals deposited as a thin film on a silicon substrate using drop-casting. Using a Lorentzian fit, the spectral data is reproduced, allowing identification of peaks linked to distinct vibrational modes within CsPbBr$_3$ nanocrystals. **d**, Quantification of up-conversion efficiency as a function of the detuning energy within the band gap. **e**, Absorption coefficient is fitted along the band tail to extract the Urbach energy, essential for anti-Stokes absorption calculations, detailed in SI.

To quantify the up-conversion efficiency, the ASPL is compared to the Stokes PL which is the cap efficiency (i.e., all excitons excited above band gap that reached the band edge). Following the work of Zhang *et al*[4], the relation between Stokes PL, ASPL, and up-conversion efficiency ($\eta$) is formulated as follows:

$$I_{spl} = I_s\, A_s\, PLQY \qquad (1)$$

$$I_{aspl} = I_{as}\, A_{as}\, \eta\, PLQY \qquad (2)$$

where $I_{spl}$ and $I_{aspl}$ are the Stokes PL and ASPL intensities, $I_s$ and $I_{as}$ are the Stokes and anti-Stokes excitation intensities, $A_s$ and $A_{as}$ are the Stokes and anti-Stokes absorption, and $\eta$ is the up-conversion efficiency. Measurement of the up-conversion efficiency (**Fig. 2a**) is performed such that $I_{spl}$ and $I_{aspl}$ are similar in value. This allows us to re-arrange equation (1) and (2) to solve for $\eta$ as follows:

$$\eta \approx \frac{I_s\, A_s}{I_{as}\, A_{as}} \qquad (3)$$

Equation (3) requires $A_{as}$ to be known for every excitation wavelength detuned into the CsPbBr$_3$ nanocrystals energy gap. Due to the low sensitivity of a typical spectrophotometer at the absorption band tail, we relied on the fact that absorption is related to the Urbach energy through the absorption coefficient. We fit the slope of the absorption coefficient from the band edge (**Fig. 2e**, more details in SI), extracted the Urbach energy, 30.95 meV, and calculated the absorption at the anti-Stokes position, $A_{as}$, at different excitation wavelengths. The Urbach energy of 30.95 meV falls within the lower range of typical Br vacancy levels, as theoretically calculated (within 100 meV), for Br-deficient CsPbBr$_3$ nanocrystals[2]. This observation is also consistent with other experimental findings, which have reported Urbach energy values ranging from 40 to 48 meV[24,25]. This suggests that our nanocrystals exhibit significantly reduced defect states. Using this Stokes and anti-Stokes data, we calculated the up-conversion efficiency for every excitation wavelength (**Fig. 2d**). These measurements yielded values for $\eta$ from 74.80% down to 3.38% for $\lambda_{exc} = 515$ nm ($\Delta E = 28.4$ meV) and $\lambda_{exc} = 550$ nm ($\Delta E = 181.6$ meV), respectively. Our results are in good agreement with previously reported values for CsPbBr$_3$ nanocrystals[5].

To further assess the role of phonons in the up-conversion process, we conducted a comparative analysis of up-conversion efficiency under two different temperature conditions: RT and liquid nitrogen (LN) temperature (~77 K). This analysis was performed on a freshly prepared sample with a PL emission peak at 2.4503 meV (as shown in Fig. S3). In both cases, we observed that the ratio of laser intensity to ASPL intensity exhibited an exponential relationship. Based on this relationship, we determined the phonon energy associated with the up-conversion process as ~ 8 meV (corresponding to 64.712 $cm^{-1}$) at RT and 6 meV (corresponding to 48.17 $cm^{-1}$) at LN temperature. Notably, these phonon energies fall within the range of the Pb-Br-Pb bending mode[7,15,16,21].

We used the Urbach energy of the new sample, 24.655 meV, to calculate up-conversion efficiencies as above. As depicted in Fig. S3f, our results clearly illustrate a significant reduction in up-conversion efficiency (~1.97-fold reduction on average) when the sample is cooled to cryogenic temperatures. Phonon population follows a Boltzmann distribution that is dependent on both energy and temperature factors. Therefore, we compared the ratio of the number of phonons ($N_P$) at RT and LN as follows:

$$\frac{N_P(RT)}{N_p(LN)} \propto \frac{e^{-\left(\frac{E_p(RT)}{\kappa_B T(RT)}\right)}}{e^{-\left(\frac{E_p(LN)}{\kappa_B T(LN)}\right)}}$$

yielding a value of 1.801. Remarkably, this value closely corresponds to the ratio of up-conversion efficiencies at the two temperature conditions (see Fig. S3f inset). We also determined that room temperature falls within the optimal temperature range for maximizing up-conversion efficiency (depicted in Fig. S4; detailed further in the SI).

To further confirm that the ASPL is coming from the band edge emission, we measured steady-state time-resolved PL at RT for above- and below-gap excitations. **Fig. 3** illustrates that the decays follow a mono-exponential form with similar lifetimes. This result suggests that, regardless of whether the electron is excited above the band edge or up-converted to the band edge, decay originates from the band edge.

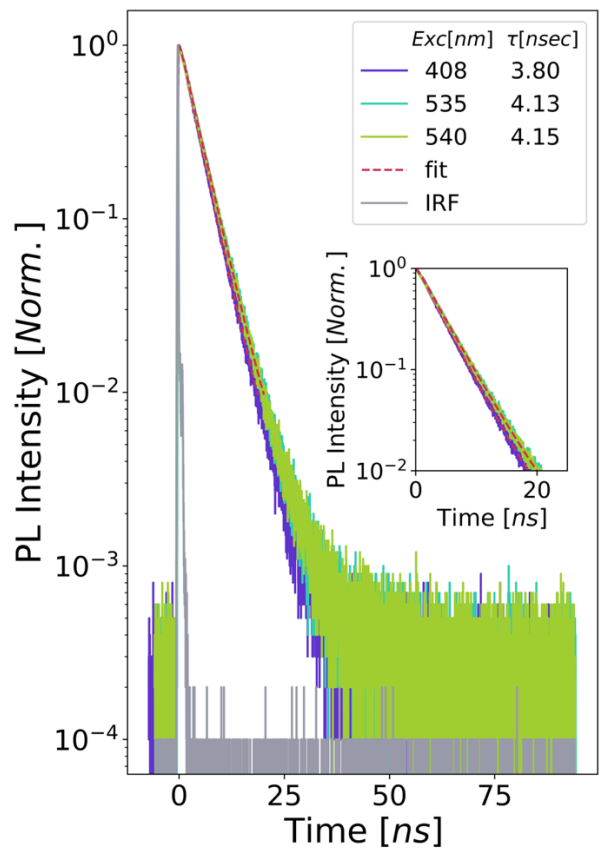

**Figure 3:** Steady-state time-resolved PL for above- and below-gap excitations. Both excitations exhibit closely matched mono-exponential decay patterns with comparable lifetimes.

**Excited State Dynamics**

To investigate how up-conversion affects electron-lattice vibration (electron-phonon coupling) in the nanocrystals, we next employed ultra-fast time-resolved spectroscopic measurements. We performed transient absorption spectroscopy (TA) to elucidate the initial dynamics of excited states within the colloidal $CsPbBr_3$ nanocrystals with above- and below-gap excitation. Above-gap excitation was achieved by delivering 31.7 $[\frac{\mu J}{cm^2}]$ of pumping pulse power with 35 fs pulses centered at 440 nm. Since intense pumping below gap can electronically excite semiconductors via two-photon absorption, we first investigated the pump fluence dependence of our below-gap excitation (at 540 nm) and followed band edge bleach signals that corresponded to electron-hole pair excitation. As presented in **Fig. 4**, a linear dependence on pump fluence clearly exists at 2 ps and 10 ps pump-probe time delays (both subsequent to typical carrier cooling timescales). We discern a linear scaling of bleach amplitude with pump power up to ~300 $[\frac{\mu J}{cm^2}]$. This linear regime indicates that signals below this power originate from a one-photon process consistent with up-conversion, helps us to avoid higher intensity regimes that would convolve up-conversion and two-photon pumping, and also allows us to judiciously select a power level that provides an optimal signal-to-noise ratio. Consequently, we set the below-gap pump power at 217.3 $[\frac{\mu J}{cm^2}]$ for this investigation (notably, Fig. S5 also demonstrates similar effects with 163 $[\frac{\mu J}{cm^2}]$).

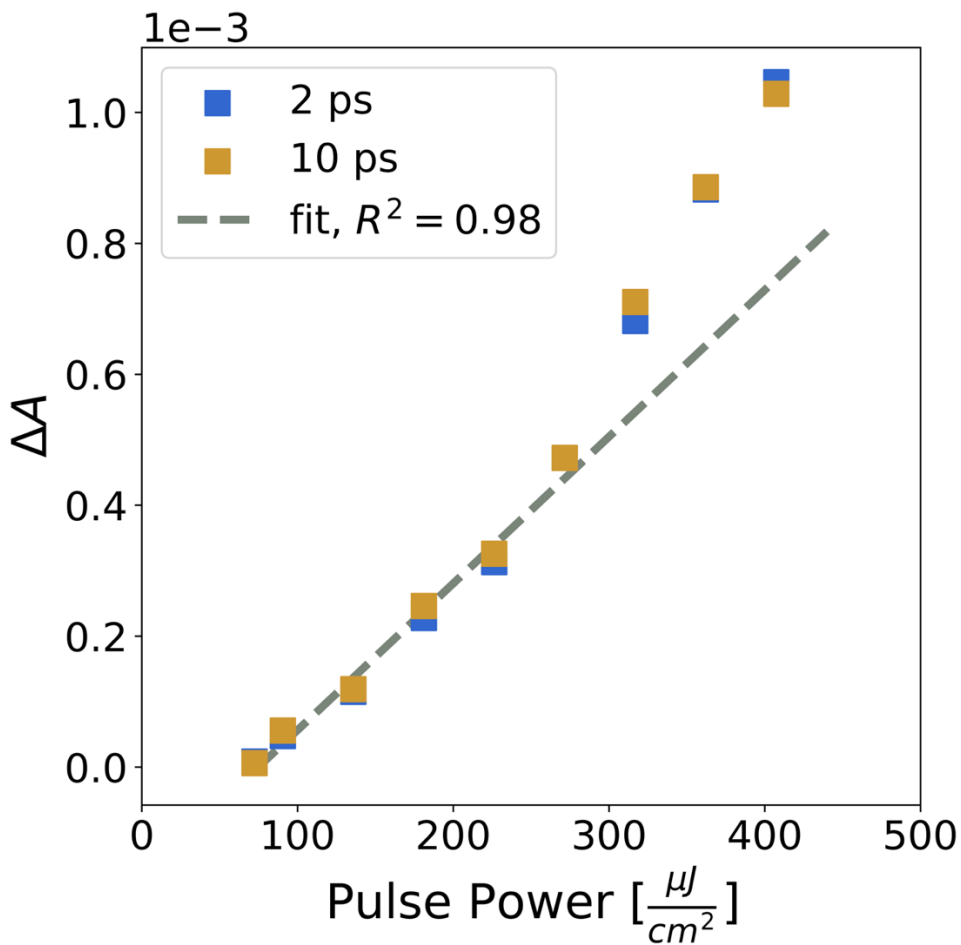

**Figure 4:** Transient absorption (TA) versus excitation power at two different timescales reveals linear and superlinear regimes. This allows to select an excitation power within the linear regime that is strong enough to yield a high signal-to-noise ratio.

For above-gap excitation (440 nm), the TA spectral map exhibits several distinct characteristics. Immediately following excitation, a pronounced induced absorption is observed near 512 nm that decays within 1 ps (**Fig. 5a, 5b**). A band-edge bleach feature also appears and completes rising within this timeframe. The red-shifted induced absorption feature is conventionally linked to the presence and relaxation of a hot exciton state where intraband cooling of electron and holes to band-edge states occurs upon phonon generation[26,27]. Within the first 1 ps, the measurements reveal a slight red-shift (~1 nm) in the exciton transition, consistent with exciton-exciton Coulomb interactions[26,28]. It is worth noting that the cooling of a hot exciton sometimes results in the generation of coherent phonons if cooling is fast compared to the phonon period, which can induce oscillatory behavior in the TA signal. Under this excitation condition here, the composition does not exhibit such a phenomenon. Upon integrating the area under the TA curve versus pump-probe time delay (**Fig. 5b** inset), a distinct plateau in the values becomes evident after 0.7 ps. This timescale aligns with polaron formation in $CsPbBr_3$ nanocrystals[7], which can also lead to induced absorption near 486 nm with a similar timescale[29] (**Fig. 5c**).

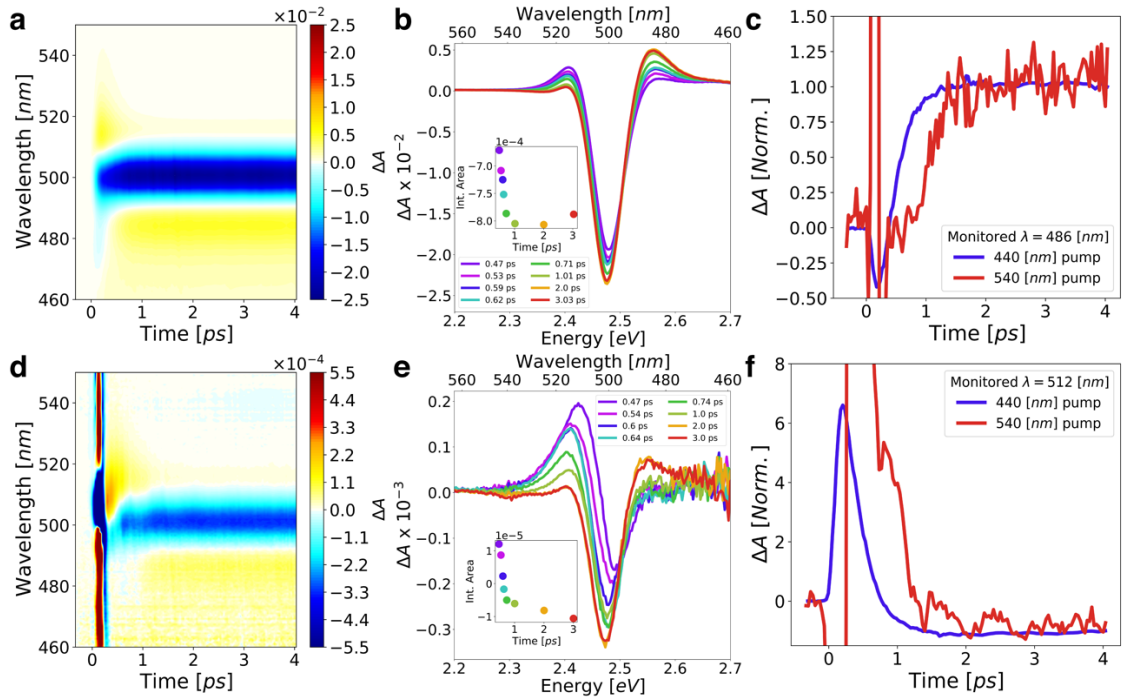

**Figure 5: a**, **d**, Time-resolved TA spectral map with (**a**) above-gap and (**d**) below-gap excitation (early time artifacts from excitation source). **b**, **e**, Selected time-traces for (**b**) above-gap and (**e**) below-gap excitation. The behavior comparison between above- and below-gap excitations reveals interesting dynamics. In the case of below-gap excitation, an initial second-derivative pattern transitions into a symmetric pattern, differing from the above-gap excitation case. Notably, within the first picosecond, the integrated area for the below-gap excitation shows a zero crossing. **c, f**, Wavelengths (**c**) above and (**f**) below the ground-state bleach are tracked over time and normalized at the final picosecond (due to the convergence of both excitations, also shown in Fig. S6). A distinct delay in the (**c**) rise and (**f**) decay of the below-gap excitation becomes evident.

For below-gap excitation (540 nm) conducted in the linear regime, distinct features appear in the TA spectra. Notwithstanding the coherent signal near time zero, a photoinduced absorption signal again emerges near 512 nm upon excitation (**Fig. 5d, 5e**) and decays on a similar timescale to the above-gap condition. Note, however, that the 512 nm induced absorption signal nominally should not appear owing to an absence of hot excitons and band-edge bleach formation should approach the instrument response function, as intraband relaxation does not occur under this condition. Time-resolved spectra also appear appreciably distinct, which we elaborate on below. Remarkably, we also observe oscillatory spectral signals emerging below the band edge in the TA spectrum in **Fig. 5d** (along the border of ground-state bleach and the photoinduced absorption), where the oscillation frequency can be quantified by performing fast Fourier transform (FFT) along the time axis. We obtained a characteristic frequency value of approximately 50 cm$^{-1}$ (illustrated in **Fig. 6a**), which corresponds to the frequency of the Pb-Br-Pb bending phonon mode. The observation of coherent phonons in this scenario implies phonons are involved in the ultrafast up-conversion and relaxation processes, which may seem counterintuitive since below-gap excitation conditions generate neither a hot electron nor a hot hole and presumably remove phonons from any excited nanocrystal. Thus, nanocrystals undergoing up-conversion contain a quantum of the specific phonon that impulsively gets removed upon absorption, and the observed oscillation reveals a dominant relaxation mechanism on the path to polaron formation. This outcome reinforces the assertion that the efficient up-conversion in CsPbBr$_3$ nanocrystals is associated with the Pb-Br-Pb bending mode, as visually represented in **Fig. 6b**.

The distinctive line shapes observed in the below-gap excitation TA spectra (**Fig. 5d**) are also worth discussing. During the initial period (<1 ps), the TA spectra exhibit characteristics of a second derivative, which subsequently transform into a symmetric above-gap-excitation-like pattern. It is important to highlight that, unlike typical above-gap excitation scenarios, the integrated area of the TA curves crosses zero as time advances. This intriguing observation suggests that the polaron generated during the early stages plays a crucial role in facilitating electron-phonon coupling for up-conversion by inducing charge redistribution and lattice distortion due to the soft lattice of CsPbBr$_3$[7,30]. The charge redistribution is marked by the noticeable red-shift (~3 nm) in the TA spectra, which can be attributed to the intricate interplay of many-body Coulomb interactions[7]. This transition in line shape occurs within a short interval of 0.7 ps, which serves as additional confirmation of the critical involvement of polarons in the up-conversion process[7]. We also observe a noticeable delay in the rise and decay of the time traces when monitoring below the band edge, as opposed to above the gap, for both above- and below-gap excitations (**Fig. 5c, 5**f).

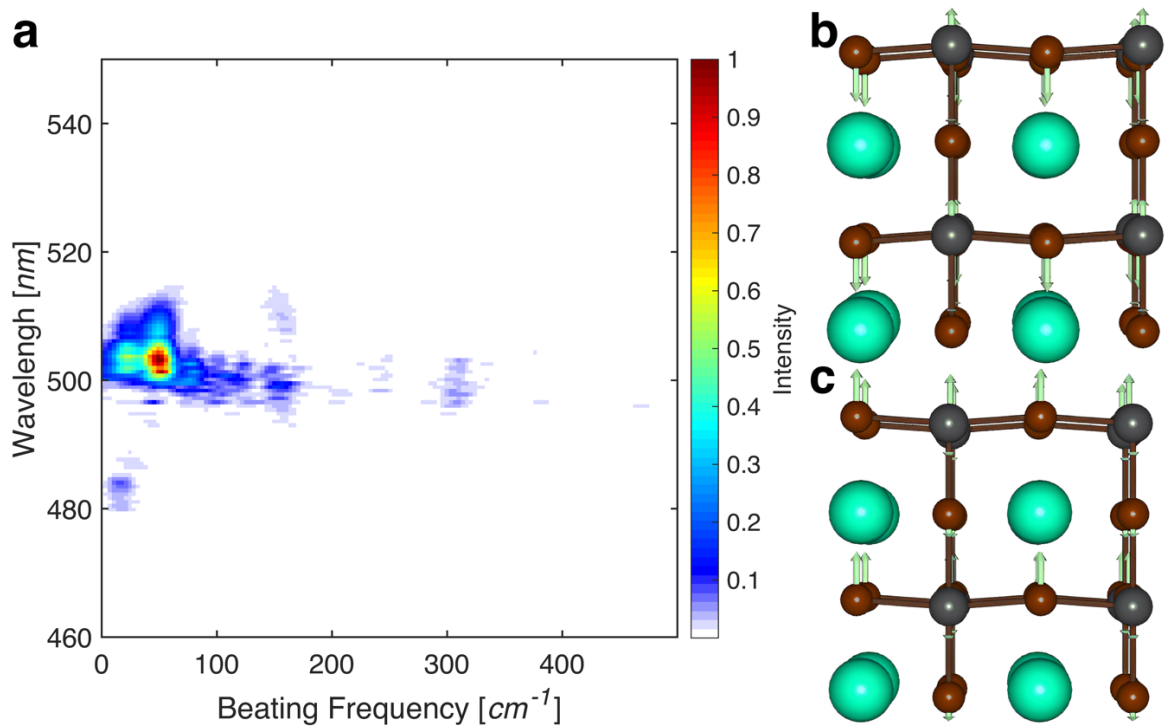

**Figure 6: a**, Beating frequency, from the oscillation components, acquired through subtraction of the exponential decay (depicted in Fig. S7). **b**, **c**, The bending mode of Pb-Br-Pb is illustrated, presenting two distinct time snapshots in panels (**b**) and (**c**).

**CONCLUSION**

In conclusion, our study on the intricate dynamics of excited states reveals that the efficient up-conversion observed in $CsPbBr_3$ nanocrystals is driven by the Pb-Br-Pb bending mode. We also reveal that, under the below-gap excitation condition, coherent phonons unexpectedly emerge at the early times within the subset of nanocrystals undergoing up-conversion as polarons form. This result indicates that the up-conversion and subsequent relaxation is accompanied by a synchronized and phased lattice motion. This study emphasizes how the intricate and unique interplay between the soft lattice structure, phonons, and excited states dynamics of $CsPbBr_3$ nanocrystals facilitate the up-conversion process. Utilization of different lead-halide compositions with dominantly higher energy phonon modes, such as the Pb-X stretching mode, may improve up-conversion efficiency.

**Data availability**

All the data supporting the findings of this study are available within this article and its Supplementary Information. Any additional information can be requested from corresponding authors.

**Acknowledgements**

This work was supported by Samsung Electronics via Samsung Advanced Institute of Technology (SAIT) under Contract No. FRA000836. Use of the GSECARS Raman Lab System was supported by the NSF MRI proposal (EAR-1531583). GeoSoilEnviroCARS was supported by the National Science Foundation – Earth Sciences (EAR – 1634415). Work performed at the Center for Nanoscale Materials, a U.S. Department of Energy Office of Science User Facility, was supported by the U.S. DOE, Office of Basic Energy Sciences, under Contract No. DE-AC02-06CH11357. The authors thank Dr. Karen M. Watters for scientific editing of the manuscript.

**Author contributions**

A.S.A. conceived this study. A.S.A. synthesized and characterized the nanocrystals and conducted steady-state experiments (excluding Raman). V.B.P and S.C. helped with Raman setup and data collection. R.D.S conducted Raman measurements. B.C.L. performed Raman background subtraction. A.S.A analyzed Raman spectra. B.C.L. and R.D.S performed and analyzed transient absorption measurements. A.S.A wrote the paper in consultation with all authors. All authors contributed to the revision of the final paper.

**Competing interests**

The authors declare no competing interests.

# Supporting Information

# Efficient Up-Conversion in CsPbBr$_3$ Nanocrystals via Phonon-Driven Exciton-Polaron Formation

*Abdullah S. Abbas[1,†], Beiye C. Li[4,6,7], Richard D. Schaller[8,9], Vitali B. Prakapenka[10], Stella Chariton[10], Gregory S. Engel[4,5,6,7], A. Paul Alivisatos[1,2,3,4,5,6, §,*]*

[1]Department of Materials Science and Engineering, University of California, Berkeley, Berkeley, California 94720, United States

[2]Department of Chemistry, University of California, Berkeley, Berkeley, California 94720, United States

[3]Materials Sciences Division, Lawrence Berkeley National Laboratory, Berkeley, California 94720, United States

[4]Department of Chemistry, The University of Chicago, Chicago, Illinois 60637, United States

[5]Pritzker School of Molecular Engineering, The University of Chicago, Chicago, Illinois 60637, United States

[6]James Franck Institute, The University of Chicago, Chicago, Illinois 60637, United States

[7]Institute for Biophysical Dynamics, The University of Chicago, Chicago, Illinois 60637, United States

[8]Department of Chemistry, Northwestern University, Evanston, Illinois 60208, United States

[9]Center for Nanoscale Materials, Argonne National Laboratory, Lemont, Illinois 60439, United States

[10]Center for Advanced Radiation Sources, The University of Chicago, Chicago, Illinois 60637, United States

Present addresses:

[†]Department of Chemistry, The University of Chicago, Chicago, Illinois 60637, United States

[§]Department of Chemistry and Pritzker School of Molecular Engineering, The University of Chicago, Chicago, Illinois 60637, United States

*To whom correspondence may be addressed: paul.alivisatos@uchicago.edu

**Urbach Energy**

One method for quantifying absorption below the band edge (referred to as anti-Stokes absorption) involves determining the Urbach energy using the absorption coefficient relationship:

$$\alpha(hv) = \alpha_o \, e^{\left(\frac{hv - E_{pl}}{E_u}\right)} \tag{1}$$

where $\alpha$ is the absorption coefficient, $hv$ is the excitation energy, $E_{pl}$ is the band edge energy, and $E_u$ is the Urbach energy.

Solving for $\alpha_o$ at the band edge is as follows:

$$I = I_o e^{-\alpha L} \tag{2}$$

$$log_{10}\left(\frac{I_o}{I}\right) = A = \alpha \, L \, log_{10}(e) \tag{3}$$

rearranging equation (3),

$$\alpha_o = \frac{A}{L \, log_{10}(e)} \tag{4}$$

where A is the absorption and L is cuvette path length (standard 1 cm). We solve for $\alpha_o$ at the band edge and find it to be $\alpha_o(A = 0.0698, \, E_{pl} = 2.436 eV) = 0.1607 \, cm^{-1}$.

Then rearranging equation (1):

$$\ln(\alpha(hv)) \propto \frac{1}{E_u} hv \tag{5}$$

We see that the Urbach energy can be extracted from the slope of the absorption coefficient as a function of excitation energy at the band tail, as shown in Figure 2e of the main text. We find the Urbach energy for our CsPbBr$_3$ nanocrystals to be $E_u = 30.95208617\ meV$

By utilizing the Urbach energy in conjunction with the absorption coefficient at the band edge, the anti-Stokes absorption can be derived from equation (3). Subsequently, this enables determination of the up-conversion efficiency.

**Temperature Dependency of Up-Conversion Efficiency**

In the main text, we examine the relationship between up-conversion efficiency and temperature. We establish that the ratio of the number of phonons ($N_p$) at two distinct temperatures, considering their respective phonon energies, directly correlates with the ratio of up-conversion efficiency at those temperatures. To investigate this connection, our initial step involves modeling the phonon energy as a function of temperature, as follows:

$$E_p(T) = E_o e^{\left(-\frac{\Delta E}{\kappa_B T}\right)}$$

We utilized the phonon energies obtained at both room temperature (RT) and liquid nitrogen (LN) temperature to determine the values $E_o$ and $\Delta E$.

The number of phonons ($N_P$) as a function of temperature follows the Boltzmann distribution:

$$N_p(T) = N_o e^{(-\frac{E_p(T)}{\kappa_B T})}$$

So,

$$\frac{N_P(RT)}{N_p(LN)} \propto \frac{e^{-\left(\frac{E_p(RT)}{\kappa_B T(RT)}\right)}}{e^{-\left(\frac{E_p(LN)}{\kappa_B T(LN)}\right)}}$$

And we know that

$$\frac{\eta(RT)}{\eta(LN)} \approx \frac{N_P(RT)}{N_p(LN)}$$

Thus,

$$\eta(T) \approx \frac{\eta(RT)}{\frac{N_P(RT)}{N_p(T)}}$$

where $N_P(RT)$ is computed directly while $\eta(RT)$ is extracted from the data in Figure 2d across various detuning energies residing below the band gap.

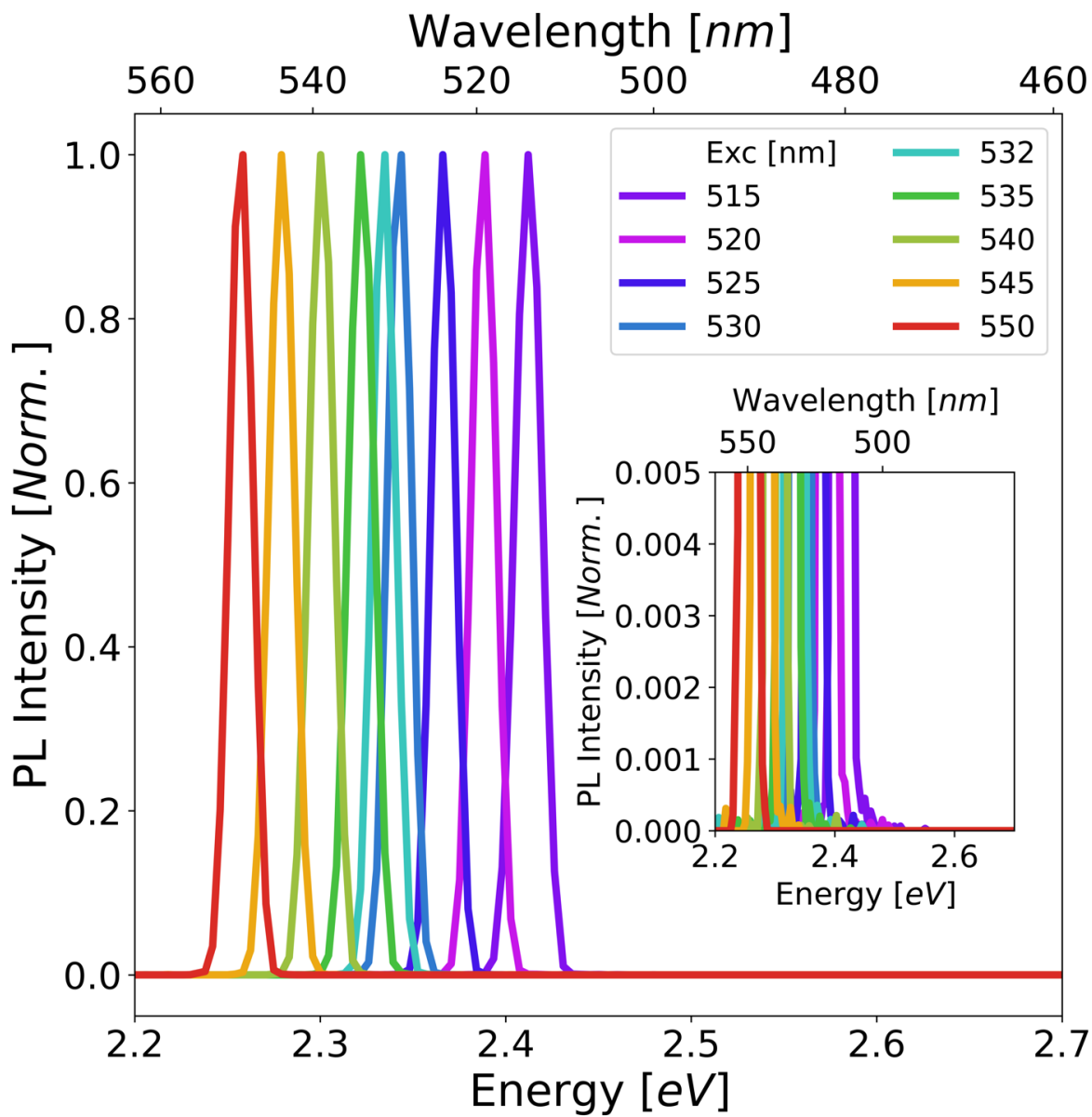

**Figure S1:** Anti-Stokes laser excitations with a 515 nm long pass filter.

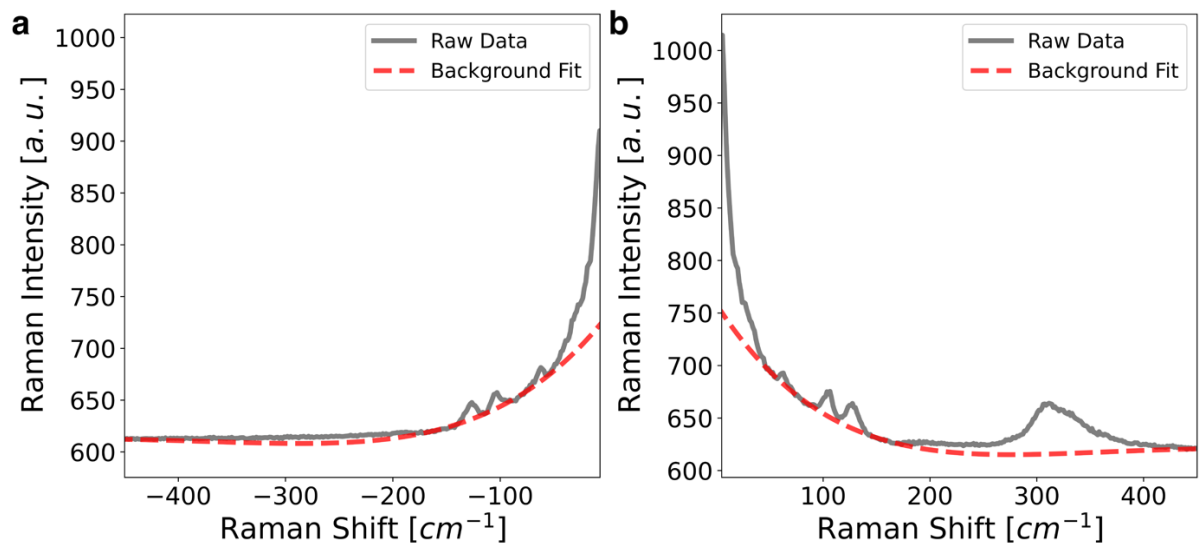

**Figure S2: a**, Anti-Stokes and **b**, Stokes Raman raw data with the background fit correction using the 5th order asymmetric truncated quadratic function.

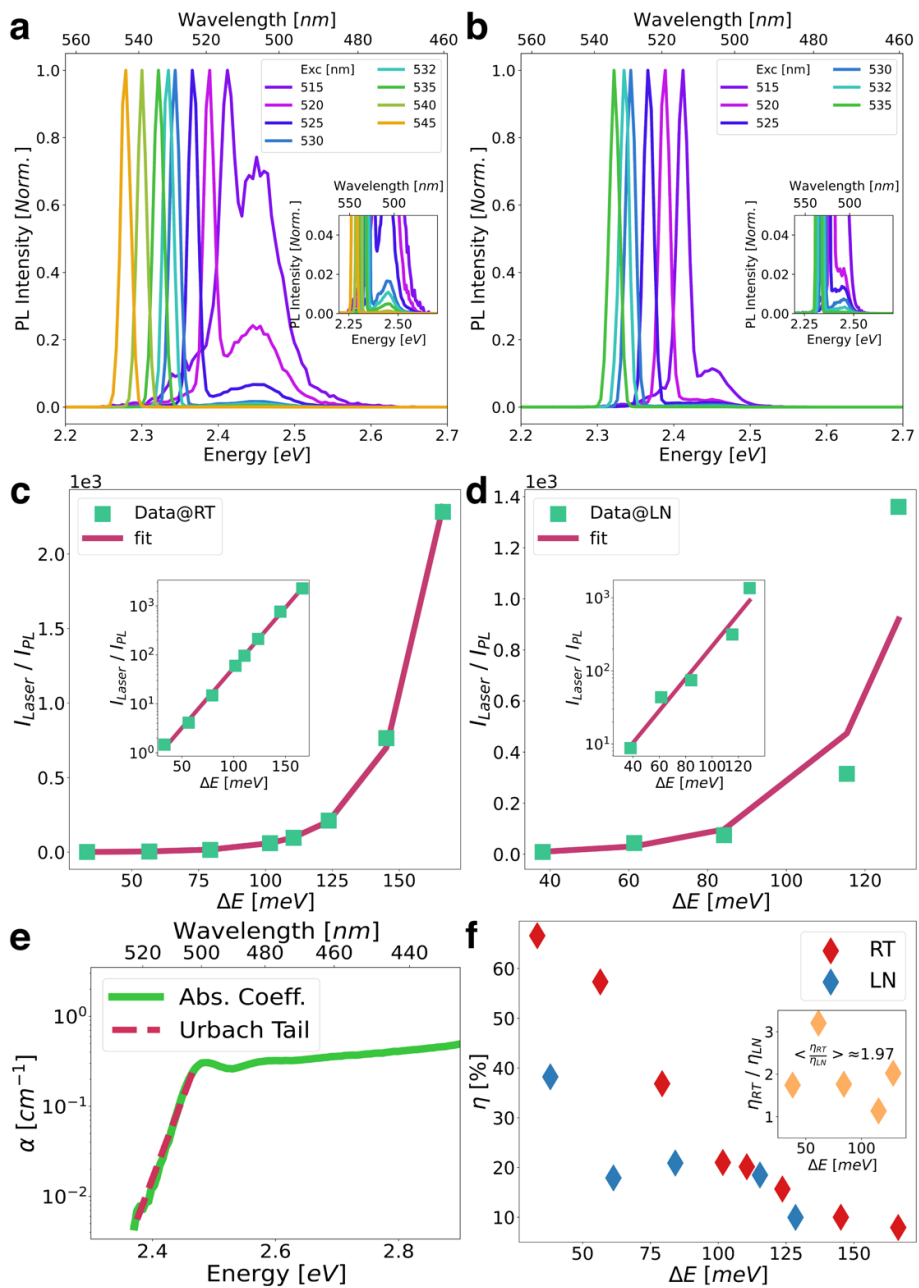

**Figure S3: Comparative Analysis of Up-Conversion Efficiency at Room Temperature (RT) and Liquid Nitrogen (LN) Temperature. a**, **b**, Below-gap excitation is illustrated for both (**a**) RT and (**b**) LN temperatures. **c**, **d**, Phonon energies of approximately 8 meV (64.712 cm$^{-1}$) and 6 meV (48.17 cm$^{-1}$) were determined by fitting the ratio of the laser intensity to the anti-Stokes

photoluminescence (ASPL) intensity curve for (**c**) RT and (**d**) LN temperatures, respectively, and fall within the Pb-Br-Pb bending mode range[1–4]. **e**, The Urbach energy of $24.655\ meV$ was obtained by fitting the tail of the absorption coefficient. **f**, Notably, the up-conversion efficiency exhibited a drastic decrease upon cooling to cryogenic LN temperatures, with a phonon population ratio closely matching the efficiency ratio.

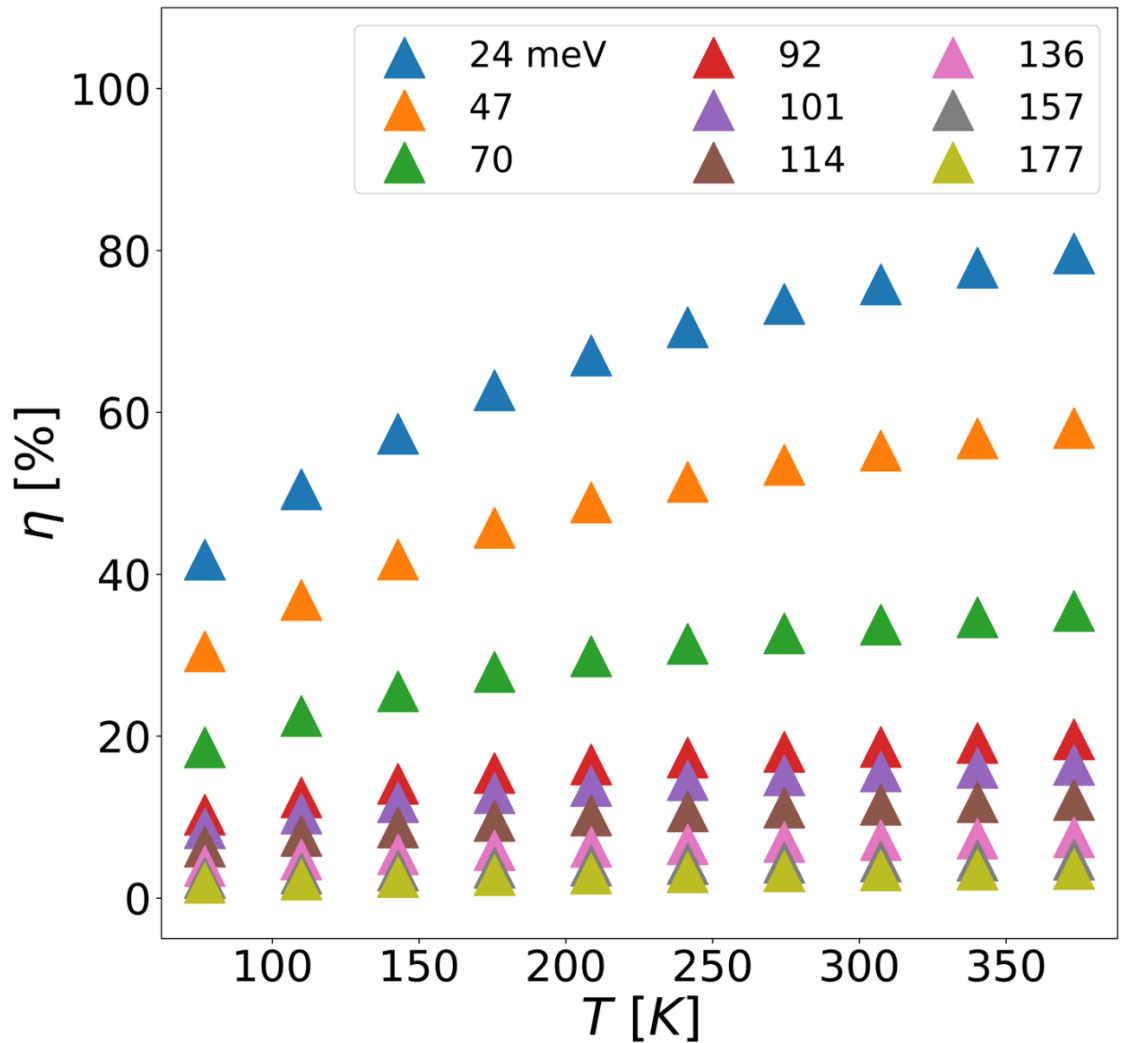

**Figure S4:** The up-conversion efficiency, as determined through the methodology described above, at various temperature conditions for different detuning energies below the band gap.

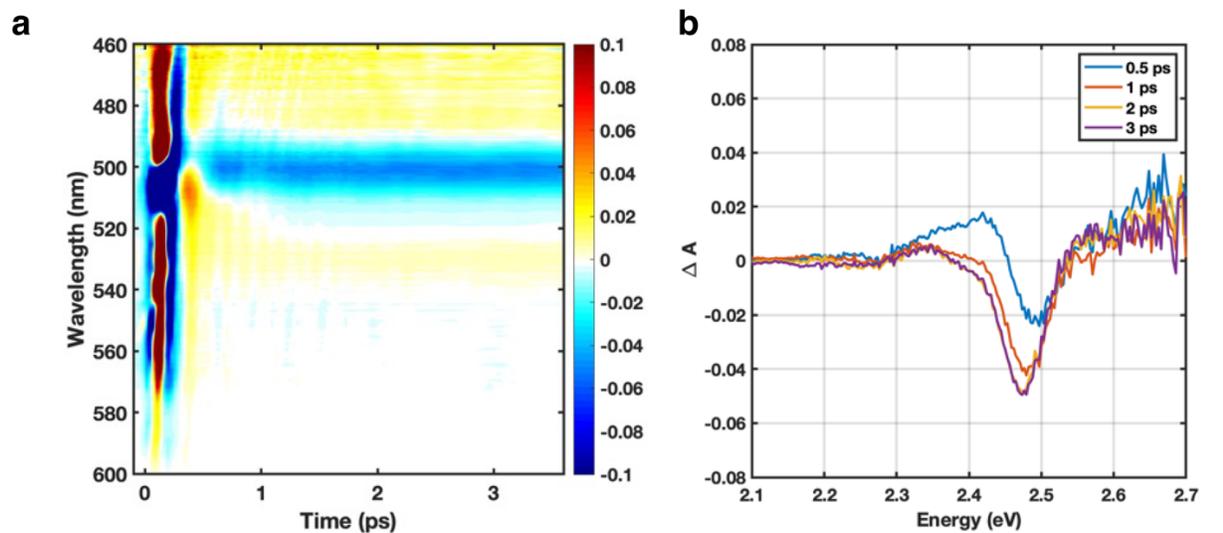

**Figure S5: a**, Time-resolved transient absorption map along with **b**, selected time traces at a low power of 163 $[\frac{\mu J}{cm^2}]$, demonstrating the persistent occurrence of coherent phonons.

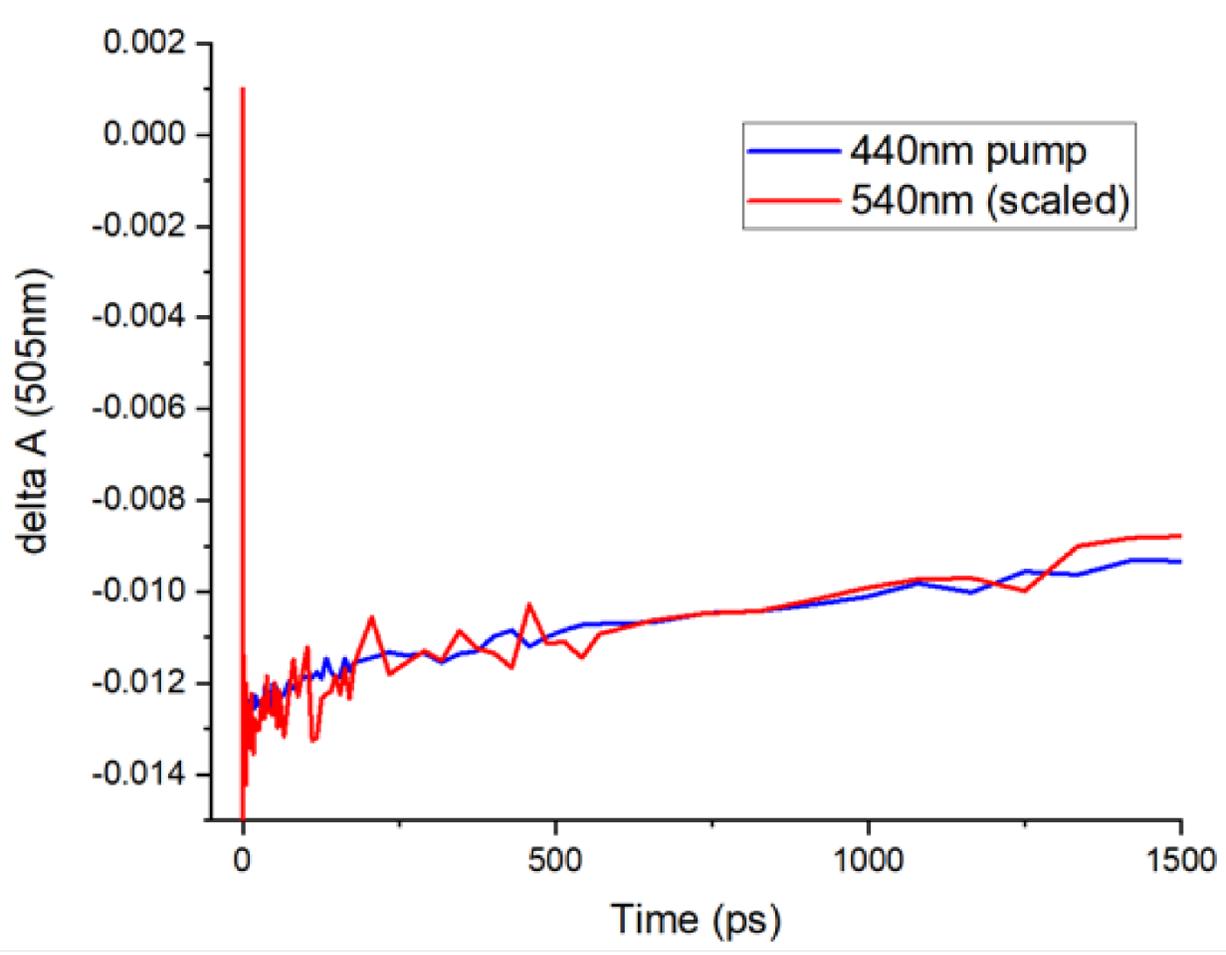

**Figure S6:** Illustration of decay traces that exhibit overlapping behavior in the long-time limit, both for resonant and non-resonant excitations.

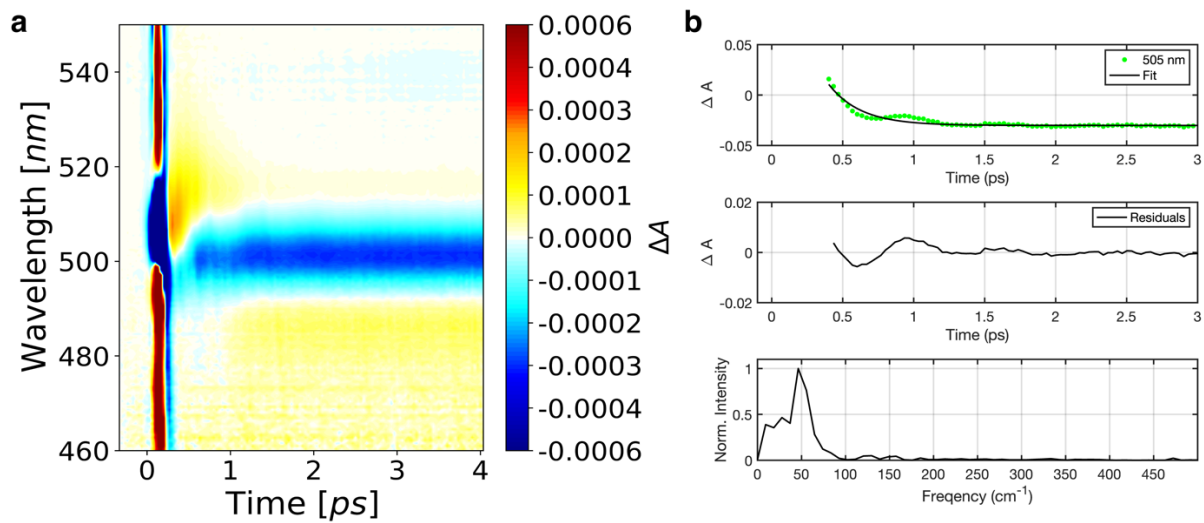

**Figure S7:** Time traces taken from **a**, time-resolved transient absorption and **b**, fitted to an exponential decay (**top panel**). The fit is subtracted to get the residual (**b-middle panel**) which is then Fourier transformed to get a beating frequency (**bottom panel**) that is attributed to a vibrational mode. This vibrational mode of ~50 cm$^{-1}$ corresponds to the Pb-Br-Pb bending mode[1–4].